\documentclass{Interspeech2024}

\interspeechcameraready 

\title{Source Tracing of Audio Deepfake Systems}

\name[]{Nicholas}{Klein}
\name[]{Tianxiang}{Chen}
\name[]{Hemlata}{Tak}
\name[]{Ricardo}{Casal}
\name[]{Elie}{Khoury}

\address{
  Pindrop, Atlanta, GA, USA
}
\email{\{nklein,tchen,Hemlata.Tak,rcasal,ekhoury\}@pindrop.com}


\usepackage{multicol,multirow}
\usepackage{url,cite}
\usepackage{hyperref}
\usepackage{colortbl}
\usepackage[dvipsnames]{xcolor}
\usepackage{tabularx,lipsum,amssymb,amsmath,pifont,graphicx,adjustbox}
\usepackage{lipsum}
\usepackage{verbatim}
\hypersetup{
    colorlinks=true,
    linkcolor=Blue,
    urlcolor=Blue,
    citecolor=BrickRed
}
\newcommand{\newpara}[1]{\vspace{8pt}\noindent\textbf{#1}}

\copyrightnotice{\footnotesize \copyright 2024 Pindrop Security, Inc. This work is licensed under \href{https://creativecommons.org/licenses/by-nc-nd/4.0/legalcode.en}{CC BY-NC-ND 4.0}.}

\begin{document}

\maketitle

\begin{abstract}
Recent progress in generative AI technology has made audio deepfakes remarkably more realistic. While current research on anti-spoofing systems primarily focuses on assessing whether a given audio sample is fake or genuine, there has been limited attention on discerning the specific techniques to create the audio deepfakes. Algorithms commonly used in audio deepfake generation, like text-to-speech (TTS) and voice conversion (VC), undergo distinct stages including input processing, acoustic modeling, and waveform generation. In this work, we introduce a system designed to classify various spoofing attributes, capturing the distinctive features of individual modules throughout the entire generation pipeline. We evaluate our system on two datasets: the ASVspoof 2019 Logical Access and the Multi-Language Audio Anti-Spoofing Dataset (MLAAD). Results from both experiments demonstrate the robustness of the system to identify the different spoofing attributes of deepfake generation systems.
\end{abstract}

\section{Introduction}
\thispagestyle{preprint}
In recent years, deepfake generation and detection have
attracted significant attention. On January 21, 2024, an advanced text-to-speech (TTS) system was used to generate fake calls to manipulate the voice of US President, Joe Biden, encouraging voters to skip the 2024 primary election in the state of New Hampshire~\cite{PM2024robocall}. This incident underscores the critical need for deepfake detection that is reliable and trusted. Thus, explainability in deepfake detection systems is crucial. Within this research area, the task of deepfake audio source attribution has recently been gaining interest~\cite{Borrelli2021SyntheticSD,Yan2022AnII,Zhang2023DistinguishingNS,zhu2022source,Yi2023ADD2T,zeng2023deepfake,tian2023deepfake,lu2023detecting,deng2024vfd}. The goal of this task is to predict the source system that generated a given utterance. For example, the study in~\cite{Borrelli2021SyntheticSD} aims to predict the specific attack systems used to produce utterances in ASVspoof 2019~\cite{wang2020asvspoof}. This approach of directly identifying the name of the system misses the opportunity to categorize the spoofing systems based on their attributes. Such attribute-based categorization allows for better generalization to spoofing algorithms that are unseen in training but are composed of building blocks, such as acoustic models or vocoders, that are seen. Along these lines, authors in~\cite{Yan2022AnII} propose a more generalizable approach by classifying the vocoder used in the spoofing system. 
Authors in \cite{Zhang2023DistinguishingNS} explore classifying both the acoustic model and vocoder, finding that the acoustic model is more challenging to predict. The work in~\cite{zhu2022source}
takes this further by proposing to classify several attributes of spoofing systems in ASVspoof 2019 LA: conversion model\footnote{In~\cite{zhu2022source}, the term ``conversion model" is used instead of ``acoustic model" to refer more generally to the encoder part of the system for both TTS and VC systems.}, speaker representation, and vocoder. However, their findings demonstrate accuracy challenges in discerning speaker representation.
Another drawback of their evaluation protocol is that the ASVspoof 2019 dataset is relatively outdated as there have been many advancements in voice cloning techniques in the last five years. Finally, their choice of categories for acoustic model and vocoder are very broad (e.g.``RNN related" for acoustic model and ``neural network" for vocoder) and may not be that useful in narrowing down the identity of the spoofing system.

\begin{figure}
    \centering
     \includegraphics[trim={6.2cm 4.88cm 3cm 3.2cm},clip,width=1.674\linewidth]{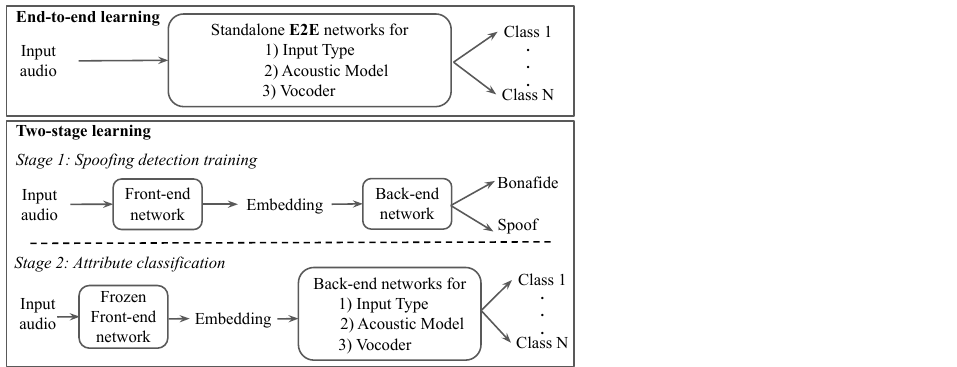}
    \caption{Illustration of proposed frameworks for spoofing attribute-classification. Top: End-to-end learning from audio. Bottom: Two-stage learning that includes a traditional countermeasure (CM) and an auxiliary classifier trained on embeddings.}
    \label{fig:frameworks}
\end{figure}

In this work, we investigate two attribute classification strategies as illustrated in Fig.~\ref{fig:frameworks}: an end-to-end learning method which trains standalone systems for each attribute and a two-stage learning method which leverages the learned representations of existing countermeasure systems. To this end, we leverage three state-of-the-art systems, namely ResNet~\cite{chen2020generalization}, self-supervised learning (SSL)~\cite{tak2022automatic}, and Whisper~\cite{kawa23b_interspeech}.
In addition to identifying the acoustic model and vocoder, we propose classifying the input type (i.e. speech, text, or bonafide) rather than speaker representation. This allows for distinguishing between TTS and VC systems. As an anchor to previous work, we evaluate our methods on the ASVspoof 2019 protocol designed by~\cite{zhu2022source}.
To address the limitations of the outdated ASVspoof-based protocol, we design a new protocol based on the recent MLAAD dataset which consists of multilingual utterances produced by 52 systems comprising a variety of state-of-the-art TTS systems. Compared to the ASVspoof-based protocol, this protocol uses more modern attack systems and replaces vague categories with specific acoustic models and vocoders. We make this novel MLAAD source tracing protocol publicly available\footnote{MLAAD protocol: \href{https://doi.org/10.5281/zenodo.11593133}{doi.org/10.5281/zenodo.11593133}}. To the best of our knowledge, this is the first study of source tracing on a multi-lingual TTS dataset. 

\section{Attribute classification of spoof systems}\label{methodology}
In this section, we describe our approaches for classifying the input type, acoustic model, and vocoder of the spoofing system used to generate a given audio.

\subsection{Proposed strategies}
We present two strategies for leveraging existing state-of-the-art (SOTA) spoofing countermeasure (CM) systems for the task of component classification:
\begin{itemize}
    \item Our \textit{End-to-End} (E2E) approach takes an existing CM architecture and trains the whole model for each of the multi-class component classification tasks separately, as depicted in the top part of Fig.~\ref{fig:frameworks}.
    \item The \textit{Two-Stage} approach, shown in the bottom of Fig.~\ref{fig:frameworks}, splits training into two steps: first an existing CM is trained for the standard binary spoof detection task; next, the CM backbone is frozen and a lightweight classification head is trained on the CM's embeddings for each separate component classification task. For the classification head, we use the simple feed forward architecture from the back-end model of the ResNet spoof detection system described in~\cite{chen2020generalization}.
\end{itemize} 
While the second approach is limited to the information that the binary-trained CM learns, it is very attractive in practice: in addition to the reduction in computational costs, existing binary systems can be trained on significantly more data than we have component labels for and enhancing them with an auxiliary head rather than replacing them with a modified E2E system is much safer for models that run in production.

\subsection{Countermeasures}
We used three different CMs to validate our hypothesis. 
These systems are well known and have reported excellent detection performance on several datasets.

\newpara{ResNet.} This system consists of a front-end spoof embedding extractor and a back-end classifier. The front-end model is known as the ResNet18-L-FM model, as detailed in~\cite{chen2020generalization,chen2021spoofprint}. To enhance the model's generalization capability, large margin cosine loss~\cite{wang2018cosface} (LMCL) and random frequency masking augmentation are applied during training. The back-end model is trained using the spoof embedding vectors for the classification tasks described in Section~\ref{methodology}. The back-end classifier is a feed forward neural network with one FC layer described in~\cite{chen2020generalization}.
\begin{table}[t]
\caption{ASVspoof 2019 LA protocol for attribute-classification tasks, adapted from~\cite{zhu2022source}.}
\vspace{-0.25cm}
\resizebox{.47\textwidth}{!}{
\begin{tabular}{ccccc}
\hline
\multicolumn{1}{|c|}{}                 & \multicolumn{1}{c|}{\textbf{\# Bonafide}} & \multicolumn{3}{c|}{\textbf{\# Spoofed}}                                                                                           \\ \hline
\multicolumn{1}{|c|}{\textbf{Sets}} & \multicolumn{1}{c|}{\textbf{-}}        & \multicolumn{1}{c|}{\textbf{Input type}} & \multicolumn{1}{c|}{\textbf{Acoustic model}} & \multicolumn{1}{c|}{\textbf{Vocoder}} \\ \hline
\multicolumn{1}{|c|}{Train}   & \multicolumn{1}{c|}{7,796}              & \multicolumn{1}{c|}{71,824}               & \multicolumn{1}{c|}{71,824}                  & \multicolumn{1}{c|}{71,824}            \\ \hline
\multicolumn{1}{|c|}{Eval}    & \multicolumn{1}{c|}{1,638}              & \multicolumn{1}{c|}{4,194}                & \multicolumn{1}{c|}{4,194}                    & \multicolumn{1}{c|}{4,194}             \\ \hline              
\end{tabular}
}
\label{tab:ASVspoof}
\end{table}

\begin{table*}[tb]
 \caption{MLAAD protocol for acoustic model classification task. Tacotron2: T}
 \vspace{-0.25cm}
 \resizebox{1\textwidth}{!}{
\begin{tabular}{|c|c|ccccccccccccc|}
\hline
        & \textbf{\# Bonafide } & \multicolumn{13}{c|}{\textbf{\# Spoofed }}\\ \hline
\textbf{Sets} & \textbf{-}                  & \multicolumn{1}{c|}{\textbf{bark}} & \multicolumn{1}{c|}{\textbf{capacitron}} & \multicolumn{1}{c|}{\textbf{fastpitch}} & \multicolumn{1}{c|}{\textbf{glowtts}} & \multicolumn{1}{c|}{\textbf{neural-hmm}} & \multicolumn{1}{c|}{\textbf{overflow}} & \multicolumn{1}{c|}{\textbf{T}} & \multicolumn{1}{c|}{\textbf{T-dca}} & \multicolumn{1}{c|}{\textbf{T-ddc}} & \multicolumn{1}{c|}{\textbf{tortoise tts}} & \multicolumn{1}{c|}{\textbf{vits}} & \multicolumn{1}{c|}{\textbf{xtts-v1}} & \textbf{xtts-v2} \\ \hline
Train   & 28,345                      & \multicolumn{1}{c|}{762}           & \multicolumn{1}{c|}{845}                 & \multicolumn{1}{c|}{859}                & \multicolumn{1}{c|}{1,866}             & \multicolumn{1}{c|}{855}                 & \multicolumn{1}{c|}{846}               & \multicolumn{1}{c|}{859}                & \multicolumn{1}{c|}{856}                    & \multicolumn{1}{c|}{2,802}                   & \multicolumn{1}{c|}{834}                   & \multicolumn{1}{c|}{15,633}         & \multicolumn{1}{c|}{4,789}             & 4,758             \\ \hline
Dev     & 6,584                       & \multicolumn{1}{c|}{159}           & \multicolumn{1}{c|}{84}                  & \multicolumn{1}{c|}{61}                 & \multicolumn{1}{c|}{1,049}             & \multicolumn{1}{c|}{65}                  & \multicolumn{1}{c|}{72}                & \multicolumn{1}{c|}{83}                 & \multicolumn{1}{c|}{72}                     & \multicolumn{1}{c|}{225}                    & \multicolumn{1}{c|}{77}                    & \multicolumn{1}{c|}{4,877}          & \multicolumn{1}{c|}{1,251}             & 1,688             \\ \hline
Eval    & 6,390                       & \multicolumn{1}{c|}{79}            & \multicolumn{1}{c|}{71}                  & \multicolumn{1}{c|}{80}                 & \multicolumn{1}{c|}{1,085}             & \multicolumn{1}{c|}{80}                  & \multicolumn{1}{c|}{82}                & \multicolumn{1}{c|}{58}                 & \multicolumn{1}{c|}{72}                     & \multicolumn{1}{c|}{973}                    & \multicolumn{1}{c|}{89}                    & \multicolumn{1}{c|}{12,490}         & \multicolumn{1}{c|}{1,960}             & 2,554             \\ \hline
\end{tabular}}
\label{tab:MLAAD_acoustic_pro}
\end{table*}

\newpara{Self-supervised learning.} SSL-based front-ends have attracted significant attention in the speech community, including spoofing and deepfake detection~\cite{jiang2020self,xie2021siamese,tak2022automatic,wang2021investigating,eom2022anti,wang2023investigating,martin2022vicomtech,wang2023spoofed}. The SSL-based CM architecture\footnote{\href{https://github.com/TakHemlata/SSL_Anti-spoofing}{github.com/TakHemlata/SSL\_Anti-spoofing}} is a combination of SSL-based front-end feature extraction and an advanced graph neural network based back-end, named AASIST~\cite{jung2022aasist}. The 160-dimensional CM embeddings are extracted prior to the final fully-connected output layer. The SSL feature extractor is a pre-trained wav2vec 2.0 model~\cite{baevski2020wav2vec,babu2021xls}, the weights of which are fine-tuned during CM training. 

\newpara{Whisper.} The Whisper model is based on an
off-the-shelf encoder-decoder Transformer architecture for automatic speech recognition (ASR)~\cite{radford2023robust}. The Whisper-based CM architecture~\cite{kawa23b_interspeech}\footnote{\href{https://github.com/piotrkawa/deepfake-whisper-features}{github.com/piotrkawa/deepfake-whisper-features}} is a combination of Whisper-based front-end feature extraction and light convolution neural network (LCNN)~\cite{wu2018light} as a back-end. For the front-end feature extraction, the Whisper embedding is concatenated with 128-dimensional linear frequency cepstral coefficients (LFCCs)~\cite{sahidullah2015comparison} along with their delta and double-delta features.  The 768-dimensional CM embeddings are extracted prior to the final fully-connected output layer. The reader is referred to~\cite{kawa23b_interspeech} for further technical details.

\section{Datasets and protocols}
Two publicly available spoofing detection benchmarks are used in our study: the ASVspoof 2019 LA~\cite{todisco2019asvspoof,wang2020asvspoof} and the most recent MLAAD dataset~\cite{muller2024mlaad}. 

\subsection{ASVspoof 2019}
The ASVspoof 2019 LA dataset has three independent partitions: train, development, and evaluation. Spoofed utterances are generated using a set of different TTS, VC, and hybrid TTS-VC algorithms~\cite{wang2020asvspoof}. To compare our methods against those presented in~\cite{zhu2022source}, we adopt their protocol partition as detailed in~Table~\ref{tab:ASVspoof}. Notably, it only includes a train and development set, so we do not do any hyper-parameter search on this protocol.
While we use the same categories as~\cite{zhu2022source} for the acoustic and vocoder tasks, we create a new ``Input type" task which is helpful to separate between TTS and VC systems.
Table~\ref{tab:ASVspoof} summarises the statistics for each partition used for the different attribute classification tasks on the ASVspoof 2019 dataset.

\begin{table}[tb]
\caption{MLAAD protocol for vocoder classification task. Multiband-mel:mul; Wavegrad:w-grad}
\vspace{-0.275cm}
\resizebox{.47\textwidth}{!}{
\begin{tabular}{|c|c|cccccc|}
\hline
\textbf{}        & \textbf{\# Bonafide} & \multicolumn{6}{c|}{\textbf{\# Spoofed}}                                                                                                                                                                                          \\ \hline
\textbf{Sets} & -                & \multicolumn{1}{c|}{\textbf{bark}} & \multicolumn{1}{c|}{\textbf{hifi-gan}} & \multicolumn{1}{c|}{\textbf{mul-gan}} & \multicolumn{1}{c|}{\textbf{univnet}} & \multicolumn{1}{c|}{\textbf{vits}} & \textbf{w-grad} \\ \hline
Train  & 28,345            & \multicolumn{1}{c|}{762}           & \multicolumn{1}{c|}{9,135}              & \multicolumn{1}{c|}{2,680}                       & \multicolumn{1}{c|}{6,473}             & \multicolumn{1}{c|}{15,633}         & 859               \\ \hline
Dev    & 6,584             & \multicolumn{1}{c|}{159}           & \multicolumn{1}{c|}{2,112}              & \multicolumn{1}{c|}{150}                        & \multicolumn{1}{c|}{1,392}             & \multicolumn{1}{c|}{4,877}          & 83                \\ \hline
Eval    & 6,390             & \multicolumn{1}{c|}{79}            & \multicolumn{1}{c|}{3,753}              & \multicolumn{1}{c|}{170}                        & \multicolumn{1}{c|}{2,135}             & \multicolumn{1}{c|}{12,490}         & 58                \\ \hline
\end{tabular}}
\label{tab:MLAAD_voco_pro}
\end{table}
\subsection{MLAAD}
MLAAD consists of TTS attacks only, however it includes 52 different state-of-the-art spoofing algorithms~\cite{muller2024mlaad}.
We manually label the acoustic models and vocoders based on the available metadata.\footnote{We use the ``model\_name" field provided in the dataset's accompanying ``meta.csv" file. System descriptions for each model\_name can be found in the Coqui-TTS~\cite{Eren_Coqui_TTS_2021} and HuggingFace repositories.}
Since MLAAD includes only TTS systems, we focus on acoustic model and vocoder classification without any input-type prediction.
For end-to-end systems such as VITS and Bark, we use the name of the full system as the acoustic model and vocoder labels. Additionally, while the MLAAD dataset labels 19 different architectures, our protocol groups several systems that are identical aside from their training data. For example, the systems ``Jenny", ``VITS", ``VITS-Neon", and ``VITS-MMS" are all labeled with the same acoustic model and vocoder category ``VITS".
For the bonafide class, we include bonafide samples from the multilingual M-AILABS dataset~\cite{mailabs}. We divide the data into train, development, and evaluation partitions while preventing speaker overlap. To enable this for the spoof samples, we assign voice labels using spherical k-means clustering on embeddings from the state-of-the-art speaker verification system, ECAPA-TDNN~\cite{desplanques2020ecapa}. We use the elbow criteria on the inertia values to select $K$=75 clusters. We remove two vocoders, Griffin-Lim~\cite{griffin1984signal} and Fullband-MelGAN~\cite{yang2021multi}, since they each have a cluster containing most of their samples. The resulting acoustic model and vocoder labels along with their number of examples in each partition are presented in Table~\ref{tab:MLAAD_acoustic_pro} and Table~\ref{tab:MLAAD_voco_pro}, respectively. \\

\section{Experimental Results}\label{experiments}

\subsection{Implementation details}
ResNet and SSL models use $4$ second (s) raw audio as input, whereas the Whisper model processes on $30s$ audio. For ResNet, LFCC features are extracted using $20ms$ window and $10ms$ frame-shift along with its delta and double delta features. Since fine-tuning large SSL models requires high GPU computation, experiments with SSL are performed with a smaller batch-size of 16 and a lower learning rate of $10^{-6}$. We used the same set-up for SSL and Whisper based models as describe in~\cite{tak2022automatic} and~\cite{kawa23b_interspeech}, respectively. SSL and Whisper based models are fine-tuned on ASVspoof and MLAAD datasets in their respective experiments, whereas the ResNet model is trained from scratch. For the auxiliary classifier, a batch size of 256 and a learning rate of $10^{-3}$ is used with no hyper-parameter tuning. The best model is chosen based on Dev set accuracy and average F1-score for ASVspoof and MLAAD experiments, respectively.

\subsection{Results on ASVspoof 2019}

\begin{table}[]
    \centering
    \footnotesize
    \setlength\tabcolsep{2pt}
    \caption{Results in terms of Accuracy (\%) on the ASVspoof 2019 LA dataset. Methods presented in \cite{zhu2022source} are included in the top two rows for comparison with our methods. We show our results when training a classification head on top of fixed embeddings from the binary CM backbone (``two-stage") as well as when training the CM backbone end-to-end for this task (``E2E").}
    \vspace{-0.265cm}
    \begin{tabular}{|c|c|c|c|}
    \hline
         \textbf{Method}&  \textbf{Input type}& \textbf{Acoustic model} & \textbf{Vocoder}\\
         \hline
         ResNet34~\cite{zhu2022source}  &  -    &  86.5    & 84.5 \\
         RawNet 2~\cite{zhu2022source}  &  -    &  88.4    & 77.5\\
         \hline
         ResNet~(two-stage)&  97.8 &  92.6  &  81.4\\
         SSL~(two-stage)&  96.7&  91.4 & 73.7\\
         Whisper~(two-stage)&  78.4 & 64.4 & 63.8 \\
         \hline
         ResNet~(E2E)&  90.5&  84.3& 83.8\\
         SSL~(E2E)&  99.9&  99.4& 84.6\\
         Whisper~(E2E)& 77.5 & 72.3 & 59.5 \\ 
         \hline
    \end{tabular}
    \label{tab:asvspoof_zhu_comparison}
\end{table}

Our results are compared with the previous study~\cite{zhu2022source} on ASVspoof 2019 in terms of unweighted accuracy in Table~\ref{tab:asvspoof_zhu_comparison}.

\newpara{Input type classification}: This study introduces a novel task, predicting input types, which the previous study did not explore. We train classification heads using fixed ResNet, SSL, and Whisper based binary spoof detection models named as, ResNet~(two-stage), SSL~(two-stage), and Whisper~(two-stage). These experiments achieve 97.8\%, 96.7\% and 78.4\% accuracy, respectively. Our SSL model fine-tuned end-to-end, SSL~(E2E), further improves accuracy to 99.9\%.

\newpara{Acoustic model classification}: Several of our models surpass the previous study's highest accuracy of 88.4\%, achieved by the multi-task-trained RawNet2 model in~\cite{zhu2022source}. Specifically, SSL~(two-stage), ResNet~(two-stage), and SSL~(E2E) achieve accuracies of 91.4\%, 92.6\%, and 99.4\% (a 12.4\% relative improvement over the previous study), respectively. The substantial increase in accuracy may be due to the fact that our models are specifically trained for these tasks, unlike the previous study's multi-task approach that jointly trained on acoustic, vocoder, and speaker representation tasks.

\newpara{Vocoder classification}: Our SSL~(E2E) model slightly outperforms the previous study with an accuracy of 84.6\% (a 0.1\% relative improvement). Unlike the acoustic model, we do not see the same level of improvement. Analyzing errors from our top-performing model, SSL~(E2E),  we find that 882 out of 896 mis-predictions occur from predicting attack A07 as ``Neural Network". Attack A07 uses a non-neural WORLD vocoder, however it also uses a GAN-based post filter that identifies areas of the waveform to mask out (See~\cite{wang2020asvspoof} for further details). This post-filter is not seen in training and must have consistently affected the final waveform in a way that mangled the resemblance to traditional vocoder audio. Aside from this one kind of error, our SSL~(E2E) model's accuracy is 99.7\%.

\begin{figure}[tb]
    \centering
     \includegraphics[trim={0.02cm 0cm 1.75cm 1.95cm},clip,width=\linewidth]{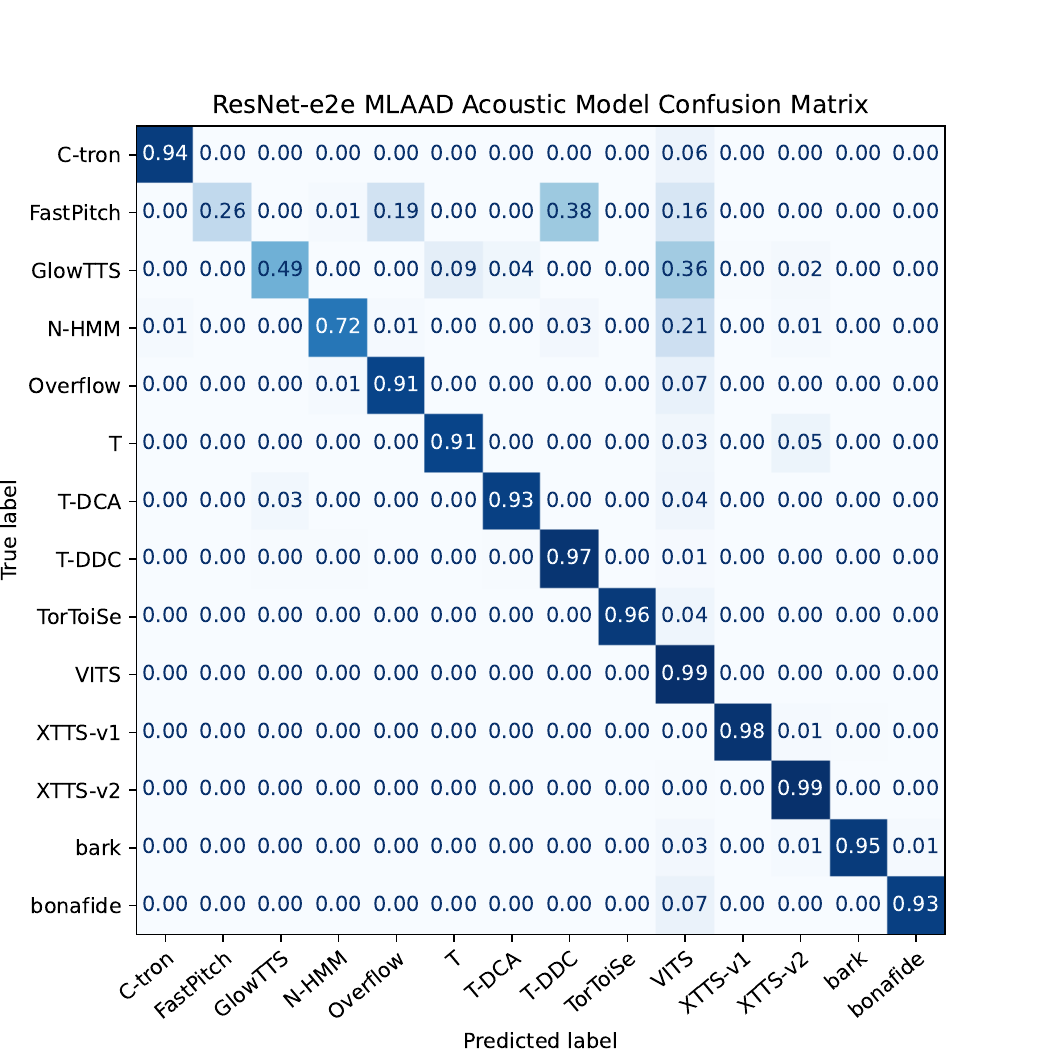}
    \caption{Confusion matrix for ResNet (E2E) acoustic model predictions on the MLAAD evaluation set. Prediction counts are normalized by true label counts (by row). T: Tacotron2}
    \label{fig:conf_mat}
\end{figure}

\subsection{Results on MLAAD}
\begin{table}
\footnotesize
    \centering
    
   \setlength\tabcolsep{7pt}
    \caption{Results in terms of macro-averaged Accuracy / F1-score (\%) on the MLAAD dataset. We show our results when training a classification head on top of fixed embeddings from the binary CM backbone (``two-stage") as well as when training the CM backbone end-to-end for this task (``E2E").}
    \vspace{-0.275cm}
    \begin{tabular}{|c|c|c|}
    \hline
         \textbf{Method}&   \textbf{Acoustic model}& \textbf{Vocoder}\\
         \hline
         ResNet~(two-stage)&       18.8 / 12.0 & 30.3 / 26.5\\
         SSL~(two-stage)&     36.6 / 16.7& 50.4 / 34.9 \\
         Whisper~(two-stage)&     49.6 / 31.5 &  48.1 / 40.2\\
         \hline
         ResNet~(E2E)&    85.4 / 82.3& 97.4 / 93.3\\
         SSL~(E2E)&  60.0 / 59.3& 93.5 / 89.4\\
         Whisper~(E2E)& 58.6 / 47.9 & 62.8 / 60.3\\
         \hline
    \end{tabular}
    \label{tab:MLAAD_res}
\end{table}

We report results in terms of macro-averaged F1 and accuracy scores in Table~\ref{tab:MLAAD_res}. With the larger number of specific vocoder and acoustic model categories compared to the ASVspoof protocol, we find that the vocoder is easier to distinguish than the acoustic model, as observed in~\cite{Zhang2023DistinguishingNS}. Our best performance on each of these tasks is achieved by our ResNet~(E2E) model, with average F1-scores of 93.3\% for the vocoder and 82.3\% for the acoustic model task. Our two-stage strategy performed noticeably worse here, indicating that the binary spoof detection models omitted much architecture-specific information when fitting to the binary task. The auxiliary head models that performed the worst on the acoustic and vocoder classification tasks are the ones that leveraged the ResNet architecture. This is likely due to the ResNet model's use of the LMCL loss function~\cite{wang2018cosface} which minimizes intra-class variation and thus reduces the separability of deepfake examples produced by different architectures.

\newpara{Error analysis:} We analyze the mistakes most commonly made by our top-performing ResNet~(E2E) model. In the acoustic model task, we get $<$90\% accuracy on three categories, as can be seen in the confusion matrix illustrated in~Fig.~\ref{fig:conf_mat}. Fastpitch is mistaken for Tacotron2-DDC 38\% of the time, Overflow 19\% of the time, and VITS 16\% of the time; GlowTTS is mistaken for VITS 36\% of the time; and Neural-HMM is mistaken for VITS 21\% of the time. In each of these cases, the predicted and actual acoustic models have a high degree of overlapping voice clusters in the test set. This indicates that the acoustic model embeddings are capturing voice information, and systems that share a common voice in the test set are more challenging to distinguish.
In the vocoder task, the ResNet~(E2E) model's performance on the different categories is high. The most mistaken category is bonafide, in which case VITS is mistakenly predicted 7\% of the time.

\begin{figure}
    \centering
     \includegraphics[trim={6.67cm 6.3cm 3cm 2.8cm},clip,width=1.37\linewidth]{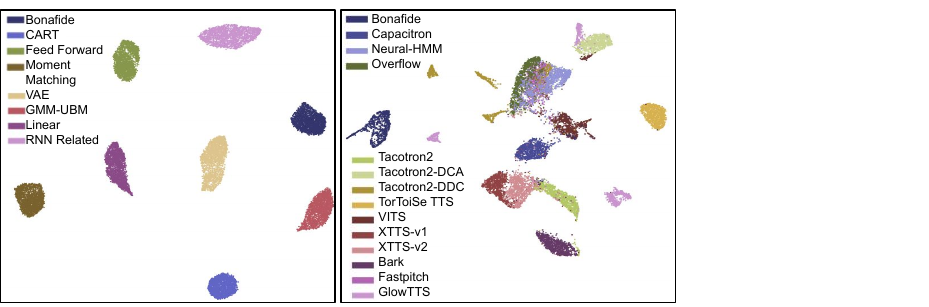}
    \caption{Embeddings from our top performing models on the acoustic model classification task of each of our protocols, plotted using UMAP dimensionality reduction with $n\_neighbors=50$. Left: ASVSpoof embeddings from SSL~(E2E) model. Right: MLAAD embeddings from ResNet~(E2E) model.}
    \label{fig:emb_vis}
\end{figure}

\subsection{Embedding visualization} 
 \label{sec:Emb_vis_exp}
Our top performing models' embeddings for the acoustic classification task using ASVspoof and MLAAD protocols are visualized using UMAP in Fig.~\ref{fig:emb_vis}. Notably, the acoustic models in the MLAAD dataset exhibit more difficulty in separation. This challenge may stem from overlapping voices among different models in the test set, as discussed in the previous error analysis section. Additionally, we observe distinct clusters of acoustic models with similar architectures: XTTS-v1 and XTTS-v2; as well as Neural-HMM~\cite{mehta2022neural} and Overflow~\cite{mehta23_interspeech} (which combines Neural-HMM with normalizing flows).

\section{Conclusions and Discussions}\label{discussion}
In this paper, we propose three multi-class classification tasks to give more explanatory predictions in the place of traditional binary spoof detection: input-type, acoustic model, and vocoder classification. We experiment with two methods of leveraging open source spoof detection systems to accomplish this task and evaluate them on a recently introduced ASVspoof 2019 protocol as well as a new protocol that we design using the more modern MLAAD dataset. Our SSL~(E2E) method outperforms the previous study on ASVspoof that we compare to on the acoustic and vocoder tasks with relative improvements in accuracy of 12.4\% and 0.1\% respectively while achieving 99.9\% accuracy on our newly introduced input-type classification task. On our MLAAD protocol which includes a greater number of vocoder and acoustic categories from more modern TTS systems, our ResNet~(E2E) model yields an average f1 score of 82.3\% for the acoustic model and 93.3\% for the vocoder classification task. Our findings support existing literature that suggest that the vocoder is easier to distinguish than the acoustic model. Additionally, we observe that the acoustic models of systems that produce similar voices are more challenging to discriminate. Thus, a potential area of future study is to more explicitly ignore voice-specific information.

Our experiments with two-stage classification methods that leverage embeddings from binary spoof detection systems show promise, though they underperform on MLAAD compared to the full model fine-tuning methods. Future research in this area is crucial as models that augment rather than replace existing binary spoof detection systems are attractive, especially in industry where changes in the behavior of the binary detection system require thorough evaluation.
Thus, one possible future experiment is to assess where in the binary model contains the most useful information for discriminating the different spoof system components. Additionally, assessing how the choice of loss function for the binary model affects the downstream multi-class performance could give insight into which existing models are best suited to being leveraged for two-stage learning.

\bibliographystyle{IEEEtran}
\bibliography{mybib}

\end{document}